\documentclass[10pt]{article}
\usepackage{setspace}
\usepackage[left = 2cm, right=2cm, top=3cm, bottom=3cm]{geometry}
\usepackage[pdftex]{graphicx}
\usepackage{cite}
\usepackage{bm}
\usepackage{float}
\usepackage{amsmath,amssymb,latexsym}
\usepackage{color}
\usepackage[T1]{fontenc}
\usepackage{wasysym}
\usepackage{siunitx}
\usepackage{setspace}
\usepackage{relsize}
\usepackage{tikz}
\usepackage{lineno, blindtext}
\usepackage{authblk}
\usepackage[symbol]{footmisc}
\usepackage{lineno}
\usepackage{upgreek}
\usepackage{pdfpages}

    \definecolor{aqua}{rgb}{0.0, 1.0, 1.0}
	\definecolor{brass}{rgb}{0.71, 0.65, 0.26}
	\definecolor{junglegreen}{rgb}{0.16, 0.67, 0.53}
	\definecolor{indigo(web)}{rgb}{0.29, 0.0, 0.51}
	\definecolor{alizarin}{rgb}{0.82, 0.1, 0.26}
	\definecolor{red(ryb)}{rgb}{1.0, 0.15, 0.07}
	\definecolor{black}{rgb}{0.0, 0.0, 0.0}
	\definecolor{darkcandyapplered}{rgb}{0.64, 0.0, 0.0}
	\definecolor{darkred}{rgb}{0.55, 0.0, 0.0}
	\definecolor{darkolivegreen}{rgb}{0.33, 0.42, 0.18}
	\definecolor{internationalkleinblue}{rgb}{0.0, 0.18, 0.65}
	\definecolor{olivedrab(web)(olivedrab3)}{rgb}{0.42, 0.56, 0.14}
	\definecolor{saddlebrown}{rgb}{0.55, 0.27, 0.07}
	\definecolor{purple(munsell)}{rgb}{0.62, 0.0, 0.77}
	\definecolor{ufogreen}{rgb}{0.24, 0.82, 0.44}
	 \definecolor{purpleheart}{rgb}{0.41, 0.21, 0.61}
		\definecolor{blue(ryb)}{rgb}{0.01, 0.2, 1.0}
			\definecolor{ao(english)}{rgb}{0.0, 0.5, 0.0}

\vspace{0.1mm}

\begin{document}
\renewcommand{\figurename}{Fig.}

\title{\color{blue}\textbf{Dichotomous behaviors of stress and dielectric relaxations in dense suspensions of swollen thermoreversible hydrogel microparticles }}
\author[1, $\dagger$]{Chandeshwar Misra}
\affil[1]{\textit{Soft Condensed Matter Group, Raman Research Institute, C. V. Raman Avenue, Sadashivanagar, Bangalore 560 080, INDIA}}
\author[2, $\ddagger$]{ Paramesh Gadige}
\affil[2]{\textit{Department of Physics, Sri Sathya Sai Institute of Higher Learning, Prasanthi Nilayam-515134, INDIA}}
\author[1,*]{Ranjini Bandyopadhyay}
\date{\today}

\footnotetext[2]{chandeshwar@rri.res.in}
\footnotetext[4]{gadigeparamesh@sssihl.edu.in}
\footnotetext[1]{Corresponding Author: Ranjini Bandyopadhyay; Phone: +91 9480836210; Email: ranjini@rri.res.in}
\maketitle
\pagebreak
\begin{abstract}
	
	{\bf Hypothesis:} While the mechanical disruption of microscopic structures in complex fluids by large shear flows has been studied extensively, the effects of applied strains on the dielectric properties of macromolecular aggregates has received far less attention. Simultaneous rheology and dielectric experiments can be employed to study the dynamics of sheared colloidal suspensions over spatiotemporal scales spanning several decades.
		\paragraph{}
		{\bf Experiments:} Using a precision impedance analyzer, we study the dielectric behavior of strongly sheared aqueous suspensions of thermoreversible hydrogel poly($N$-isopropylacrylamide) (PNIPAM) particles at different temperatures. We also perform stress relaxation experiments to uncover the influence of large deformations on the bulk mechanical moduli of these suspensions. 
		
		\paragraph{}
		{\bf Findings:} The real parts of the complex dielectric permittivities of all the sheared PNIPAM suspensions exhibit distinct relaxation processes in the low and high frequency regimes. At a temperature below the lower consolute solution temperature (LCST), both real and imaginary parts of the permittivities of highly dense PNIPAM suspensions decrease with increase in applied oscillatory strain amplitudes. Simultaneously, we note a counter-intuitive slowdown of the dielectric relaxation dynamics. Contrary to our rheo-dielectric findings, our bulk rheology experiments, performed under identical conditions, reveal shear-thinning dynamics with increasing strain amplitudes. We propose the shear-induced rupture of fragile clusters of swollen PNIPAM particles to explain our observations.
		
		\paragraph{}	
		
		\paragraph{}	
		{\bf Abbreviations :} LCST, Lower consolute solution temperature; PNIPAM, Poly($N$-isopropylacrylamide); ERFs, Electrorheological fluids; DLS, Dynamic light scattering; DSC, Differential scanning calorimetry; LAOS, Large amplitude oscillatory strains; RPM, Revolutions per minute; SI, Supporting Information; EP, Electrode polarization. 
\end{abstract}
\pagebreak
\section{Introduction}
Soft materials such as colloidal suspensions, gels, glasses, polymers and biological macromolecules can deform easily under the influence of thermal stresses at room temperature. When subjected to large external shear stresses and strains, they display fascinating flow properties over a hierarchy of spatiotemporal scales \cite{dielctricsoftmaterials}. The application of appropriate shear profiles can cause build-up or disruption of fragile suspension structures, in phenomena that are commonly referred to as shear-induced thickening or thinning respectively. It would be of interest to systematically study how a material's microscopic properties dictate its bulk flow behavior. Dielectric spectroscopy, which probes how an externally applied oscillatory electric field interacts with a material’s permanent or field-induced dipoles at different oscillatory frequencies, is a widely used technique to study complex metastable phenomena such as the dynamical slowing down process in glass-forming liquids \cite{T_Pakula_1998,T_Nicolai_1998}. The rheo-dielectric technique, which involves the study of the dielectric response of a material in the presence of shear forces, has been used to directly determine how shear-induced structural changes alter charge distributions and polarization fluctuations in nematic liquid crystals, polymers, electrorheological fluids (ERFs), viscoplastic droplets and elastomers \cite{CAPACCIOLI20074267,HWATANABE1999,Peng_Yiyan_2005,watanable2003,KNAPIKKOWALCZUK2020112494, Negita_PhysRevE.80.011705,Horio2014,Pplacke1995,B_Nath_2020,Steinhauser_2016}. Simultaneous measurements of dielectric and viscoelastic properties can therefore help us to correlate material properties over length and time scales spanning several decades.
      
\paragraph{}
Thermoresponsive poly($N$-isopropylacrylamide) (PNIPAM) hydrogel particles when dispersed in water show swelling and deswelling behaviors respectively below and above the lower consolute solution temperature (LCST, also referred to as the volume phase transition temperature, VPTT) of $\approx$ 34$^{\circ}$C \cite{HeskinsM1968,Hirokawa1984}. Below the LCST, the hydrophilic amide groups of PNIPAM form hydrogen bonds with water molecules such that individual PNIPAM particles absorb water and swell significantly. Due to increased entropy above the LCST, water molecules reorient around the non-polar regions of the PNIPAM macromolecules, thereby restricting hydrogen bond formation with the constituent amide groups. PNIPAM particles, therefore, exhibit hydrophobic behavior above their LCST and shrink by expelling water \cite{SCHILD1992}. As a consequence of this thermoreversible swelling behavior, the volume fraction of PNIPAM particles in aqueous suspensions can be controlled simply by altering the temperature of the medium at a fixed particle concentration \cite{Appel2016}. This offers an advantage over many temperature-insensitive hard colloidal particles such as poly(methyl methacrylate) (PMMA) and polystyrene (PS). Dense aqueous suspensions of PNIPAM particles have therefore been recognized as excellent model colloidal systems in the study of glass transition dynamics \cite{Yunker2014,SanjayPRM}. Various protocols, for example, synthesizing the particles in different ionic liquids or changing the concentration and nature of crosslinkers during synthesis have been implemented to alter the LCST of these particles \cite{K_jain_2015}. Besides being thermoresponsive, PNIPAM particles are deformable and compressible in aqueous suspensions at high particle volume fractions and a fixed temperature. Given their compressibility and deformability, these particles can be packed well above the random close packing fraction, $\phi_{rcp}$ = 0.64, of monodisperse hard spheres. The unique characteristics of PNIPAM particles have led to their use as drug delivery agents \cite{DONG1991,nolan2005} and biosensors \cite{RIslam2014}. 
\paragraph{}
In the existing scientific literature, researchers have independently investigated the temperature dependent phase behavior of aqueous PNIPAM suspensions using rheological and dielectric measurements \cite{Senff1999,CMisra2020,J_Wu_2003,Wu_2003_PRL}. Romeo et al. have reported $via$ bulk rheology experiments that PNIPAM particles in dense aqueous suspensions undergo a glass-liquid-gel transition with increase in temperature across the LCST due to changes in the Flory-Huggins parameter \cite{Romeo2010}. Using UV-visible spectroscopy and thermodynamic perturbation theory, Wu et al. \cite{Wu_2003_PRL} have investigated the temperature dependent phase behavior of aqueous PNIPAM suspensions. While PNIPAM particles were seen to form crystalline structures below the LCST, the suspension liquefied at a temperature near the LCST and phase separated at higher temperatures. Su et al. \cite{Wenjuan2014} have systematically studied the dielectric behavior of suspensions of PNIPAM particles of different crosslinker distributions (dense core (DC), loose core (LC) and homogeneously (HOMO) cross-linked particles). While the LCST does not change very much, the dielectric parameters of the aqueous PNIPAM suspensions near the LCST display significant changes with change in the distribution of crosslinkers. Dielectric spectroscopy experiments performed by Young et al. \cite{ManYang2017} have confirmed that dense PNIPAM suspensions exhibit a colloidal crystal to liquid transition upon increasing temperature below the LCST. While most of the previous experimental studies focused exclusively on the rheological or dielectric properties of aqueous PNIPAM suspensions \cite{Wenjuan2014,ManYang2017,J_Zhou2012}, rheo-dielectric studies investigating the dielectric response of colloidal hydrogel suspensions under applied strains have never been reported to the best of our knowledge. 
\paragraph{}
We attempt to bridge this gap by performing rheo-dielectric experiments to study the dielectric response of PNIPAM colloidal suspensions under large applied deformations for different particle effective volume fractions and solvent temperatures. We synthesize PNIPAM particles using a free radical precipitation polymerization procedure and characterize the thermoresponsive behavior and LCST of these particles in aqueous suspensions using dynamic light scattering (DLS) and differential scanning calorimetry (DSC). In the rheo-dielectric measurements, we apply large amplitude oscillatory strains (LAOS) to the sample while simultaneously measuring their dielectric responses using a precision impedance analyzer. We analyze the real and imaginary parts of the dielectric permittivities of aqueous PNIPAM suspensions of different effective volume fractions, subjected to large applied strain amplitudes, at temperatures below, near and above the LCST over a wide frequency range of the applied electrical voltage. For all the samples investigated in this study, the real parts of the dielectric permittivities display two distinct dielectric relaxation processes in the low and high frequency regimes. We extract the dielectric relaxation strengths and relaxation times at different strain amplitudes by fitting the experimental data to the Cole-Davidson relaxation model \cite{Koji2002}. In earlier studies, a low frequency dielectric relaxation process was attributed to segmental motion of the polymers and the motion of counterions along the polymer chain over the entire temperature range \cite{ManYang2017}. The high frequency dielectric relaxation was understood to originate from counterion fluctuations and interfacial polarization below and above the LCST respectively \cite{ManYang2017}.
\paragraph{}
For densely-packed aqueous PNIPAM suspensions below the LCST, we report that both the real and imaginary contributions of the dielectric permittivities decrease with increase in strain amplitudes. In contrast, the dielectric responses of densely-packed PNIPAM suspensions at temperatures near and above the LCST, and of loosely-packed suspensions at all experimental temperatures, are insensitive to the applied strain. When the applied strain amplitude is increased systematically, we observe significant increases of the dielectric relaxation times of densely-packed suspensions at a temperature below the LCST. This indicates the presence of a counter-intuitive dynamical slowing down process at very short time scales. Interestingly, we also observe a simultaneous speeding up of the stress relaxation process in bulk rheological measurements. Therefore, while our dielectric relaxation data indicates dynamical slowing down at the shortest accessible length scales, our bulk rheological data points to simultaneous shear thinning rheology at macroscopic length scales. We attribute the observed length scale dependent response functions in densely-packed suspensions below the LCST to the shear-induced formation and rupture of clusters of swollen PNIPAM particles. We propose that while the restricted motion of the entangled PNIPAM chains under shear increases the dielectric relaxation time, the shear-induced disruption of the swollen PNIPAM microstructures accelerates the stress relaxation in the bulk. Our study, therefore, provides insights into the dynamics of PNIPAM particles in aqueous suspensions over several decades of length and time scales. Such length scale dependent rheology of sheared densely-packed PNIPAM suspensions can have far-reaching implications in the use of colloidal hydrogels in electrical energy storage devices such as flow batteries \cite{Helal_2016} and flow capacitors \cite{Presser_2012}.
 
 \section{Material and methods} 
\subsection{Synthesis of poly($N$-isopropylacrylamide) (PNIPAM) particles}
PNIPAM particles were synthesized by following the one-pot free radical precipitation polymerization method. All the chemicals were purchased from Sigma-Aldrich and used as received without further purification. In the polymerization reaction, 7.0g $N$-isopropylacrylamide (NIPAM) (>99\%), 0.875g $N,N’$-methylenebisacrylamide (MBA) (>99.5\%) and 0.03g sodium dodecyl sulphate (SDS) were dissolved in 470ml Milli-Q water (Millipore Corporation, resistivity 18.2 M$\Omega$ cm) in a three-necked round bottomed (RB) flask attached with a reflux condenser, a magnetic stirrer with heating (Heidolph), a platinum sensor and a nitrogen gas (N${_2}$) inlet/outlet. The solution was stirred at 600 RPM and purged with N${_2}$ gas for 30min to remove the oxygen dissolved in water. The free radical precipitation polymerization reaction was initiated by the addition of 0.28g of potassium persulphate (KPS) (99.9\%) dissolved in 30ml Milli-Q water after heating the mixture to 70$^{\circ}$C. The reaction was allowed to proceed for 4h with a constant stirring speed of 600 RPM. After 4h, the suspension was cooled down to room temperature. Four successive centrifugations and re-dispersions at a rotational speed of 20,000 RPM for 60min were used to purify the suspension. After centrifugation, the supernatant was filtered out and the remaining sample was dried by evaporating the water. A fine powder was prepared by grinding the dried particles using a mortar and pestle. 
 
\subsection{Sample preparation and characterization of PNIPAM particles}
Aqueous PNIPAM suspensions of desired concentrations/effective volume fractions were prepared by stirring PNIPAM powder in Milli-Q water. PNIPAM particles are soft and can deform when packed above the random close packing volume fraction of monodisperse hard spheres. Due to their deformability, the volume fraction $\phi$ is not the appropriate parameter to quantify the packing of these particles in suspension. Therefore, we use a modified parameter called the effective volume fraction, $\phi_{eff}$, as a quantitative measure of the extent of particle packing in suspension. The protocol for computing $\phi_{eff}$ for suspensions of PNIPAM particles in water is discussed in detail in section 1 of Supporting Information (SI). The average sizes and the thermoreversible swelling-deswelling transition of PNIPAM particles in a dilute aqueous suspension were recorded using a Brookhaven Instruments Corporation (BIC) BI-200SM dynamic light scattering (DLS) setup attached with a 150mW solid-state laser (NdYVO${_4}$, Coherent Inc., Spectra Physics) having an emission wavelength of 532nm. We observe that the average hydrodynamic diameter of PNIPAM decreases abruptly when temperature is raised (Fig. 1(a)), thereby signalling a swelling-deswelling transition of the hydrogel particles. We note that the thermoreversible PNIPAM particles swell maximally below 20$^{\circ}$C and collapse fully above 45$^{\circ}$C. The lower consolute solution temperature (LCST) of the aqueous PNIPAM suspensions was quantified using differential scanning calorimetry (DSC) (Mettler Toledo, DSC 3). The temperature corresponding to the endothermic peak ($\approx$ 34.3$^{\circ}$C) in Fig. S2 of SI represents the LCST of the PNIPAM particles in their aqueous suspensions.
\begin{figure}[!t]
	\centering
	\includegraphics[width=3in,keepaspectratio]{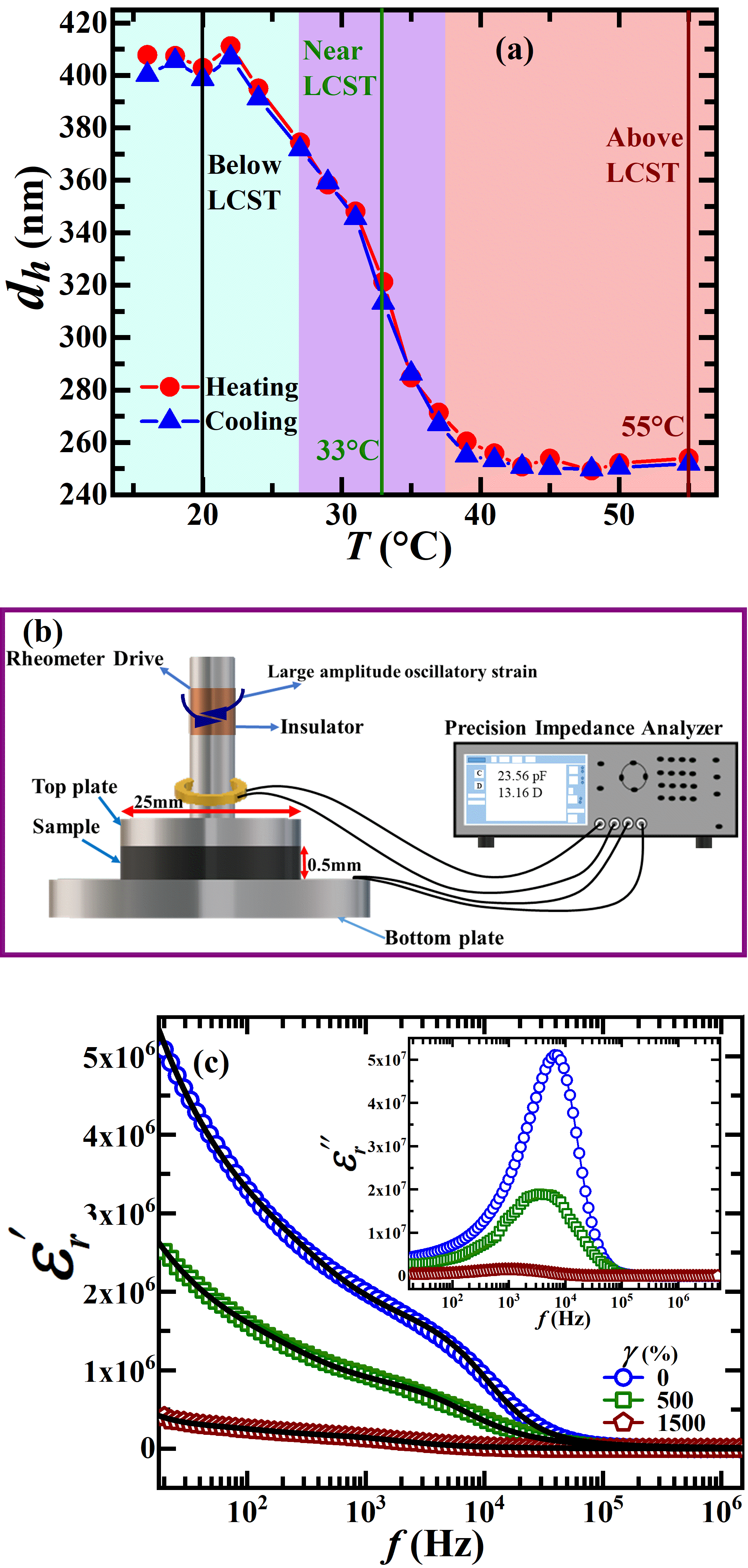}
	\caption{(a) DLS measurements of the hydrodynamic diameters of thermoreversible PNIPAM particles during heating and cooling temperature cycles. The different background colors represent the three different regimes showing the distinct temperature dependent responses of the thermoreversible PNIPAM; from left to right: below LCST, near LCST and above LCST, respectively. Measurements were made at 20$^{\circ}$C, 33$^{\circ}$C and 55$^{\circ}$C, which are marked with vertical lines. (b) Schematic representation of a rheo-dielectric setup. (c) Dependence of the real part of the relative dielectric permittivity ($\varepsilon_{r}^{'}$) of an aqueous PNIPAM suspension of volume fraction $\phi_{eff}$ = 1.9 on the frequency of the AC electrical voltage for various applied strain amplitudes at 20$^{\circ}$C (below the LCST). The solid lines are fits to the real part of Eqn. 1. The inset shows the imaginary part of the dielectric permittivity ($\varepsilon_{r}^{''}$) of an aqueous PNIPAM suspension of volume fraction $\phi_{eff}$ = 1.9 for the same applied strain amplitudes at 20$^{\circ}$C.}
	\label{Fig.1}
\end{figure}
\subsection{Rheo-dielectric measurements}
Dielectric measurements of aqueous PNIPAM suspensions were performed using a precision impedance analyzer (Wayne Kerr Electronics, 6500B series). An alternating electrical voltage of amplitude 750mV over a frequency range between 20Hz and 5MHz was applied, while a stress-controlled Anton Paar MCR 702 rheometer was used to simultaneously apply oscillatory shears to the sample. A parallel plate sample geometry (PP-25/DI/TI) with a diameter of 25mm and a measuring gap of 0.5mm was used. The sample was sandwiched between two plates that simultaneously served as capacitor electrodes and shearing geometry. Fig. 1(b) displays a schematic diagram of the rheo-dielectric setup. Oscillatory shear strains of amplitudes between 10\% to 2000\% at a fixed angular frequency of 10rad/sec were applied to study the dielectric properties of aqueous PNIPAM suspensions under flow. The impedance analyzer was connected to the electrodes (both top and bottom plates) with shielded wires. The measured capacitance ($C$) and loss factor ($D$) of aqueous PNIPAM suspensions were used to estimate the real and imaginary parts of the relative dielectric permittivities ($\varepsilon_{r}^{'}$ and $\varepsilon_{r}^{''}$) using the equations: $\varepsilon_{r}^{'} = C(d/A\varepsilon_{0})$ and $\varepsilon_{r}^{''} = D\varepsilon_{r}^{'}$, where $d$, $A$ and $\varepsilon_{0}$ are the gap between the two electrode plates (0.5mm in these experiments), the cross-sectional area of the top plate and the permittivity of free space respectively. Representative plots of the real and imaginary parts of the dielectric permittivities, measured for a PNIPAM suspension of effective volume fraction 1.9 below the LCST under different oscillatory strains, are displayed in Fig. 1(c). In observations that are consistent with previous studies \cite{Wenjuan2014, ManYang2017}, we note that the real part of the dielectric permittivity, $\varepsilon_{r}^{'}$, exhibits two distinct relaxation processes in the low and high frequency regimes. As discussed earlier in the context of PNIPAM suspensions \cite{ManYang2017}, the relaxation observed at frequencies below approximately 1kHz (the low frequency relaxation) can be attributed to the local segmental motion of the polymer and also to the motion of counterions along the polymer chain over the entire temperature range. Fig. S3(a) of SI represents a schematic representation of the segmental motion of a PNIPAM chain. 
\paragraph{}
The relaxation process that we observe at frequencies greater than 1kHz (high frequency dielectric relaxation) in Fig. 1(c) can be attributed to the fluctuations of counterions (due to the motion of Na$^{+}$ counterions around the SO$_{4}^{-}$ groups dissociated from sodium dodecyl sulphate (SDS) used during PNIPAM synthesis) below the LCST and to interfacial polarization (schematic illustration displayed in Fig. S3(b)) above the LCST \cite{ManYang2017}. Since PNIPAM particles swell fully by absorbing water at a temperature below the LCST, a clear interface between PNIPAM particles and water does not exist. Therefore, the main contribution to the fast dielectric response below the LCST arises from the electric field-induced displacement and rearrangement of counterions bound to the charged colloidal particles \cite{ManYang2017}. In contrast, PNIPAM particles collapse and expel most of the absorbed water above the LCST, thereby revealing a clear particle-water interface. Hence, at a temperature above the LCST, interfacial polarization (the separation of counterion charges at the interface) dominates the high frequency relaxation process. The acquired dielectric data has tiny fluctuations presumably arising due to the presence of metallic electrical contacts \cite{watanable2003}. In the present work, the raw dielectric permittivity data have been smoothed by averaging over 10 adjacent data points (Fig. S4 of SI).

The acquired frequency dependent dielectric permittivity data was fitted to the Cole-Davidson relaxation model by incorporating two relaxation terms that correspond to the low and high frequency regimes (the slow and fast relaxation processes respectively), while also accounting for electrode polarization \cite{Wenjuan2014, ManYang2017}:

\begin{equation}
    \varepsilon_{r}^{*}=\varepsilon_{h}+\frac{\varepsilon_{l}-\varepsilon_{m}}{1+(j\omega\uptau_{s})^{\beta_{s}}}+\frac{\varepsilon_{m}-\varepsilon_{h}}{1+(j\omega\uptau_{f})^{\beta_{f}}}+A\omega^{-m}
\end{equation}
Here, $\varepsilon_{l}, \varepsilon_{m}$ and $\varepsilon_{h}$ are the dielectric permittivities in the low, intermediate and high frequency regimes respectively, $\uptau_{s}$ and $\uptau_{f}$ are respectively the slow and fast dielectric relaxation times, while $\beta_{s}$ and $\beta_{f}$ are the corresponding exponents which vary between 0 to 1. It is to be noted that $\Delta\varepsilon_{l}$ (=$\varepsilon_{l}-\varepsilon_{m}$) and $\Delta\varepsilon_{h}$ (=$\varepsilon_{m}-\varepsilon_{h}$) are the dielectric relaxation strengths in the low and high frequency regimes respectively. The term $A\omega^{-m}$, where $A$ and $m$ are adjustable parameters, is introduced to account for electrode polarization (EP) which occurs at lower frequencies due to the accumulation of free charges at the electrode interface. These charged species presumably originate from the electrolyte (potassium persulfate KPS) that we employed as a reaction initiator during the synthesis process. 
\begin{figure}[!t]
	\centering
	\includegraphics[width=6.3in,keepaspectratio]{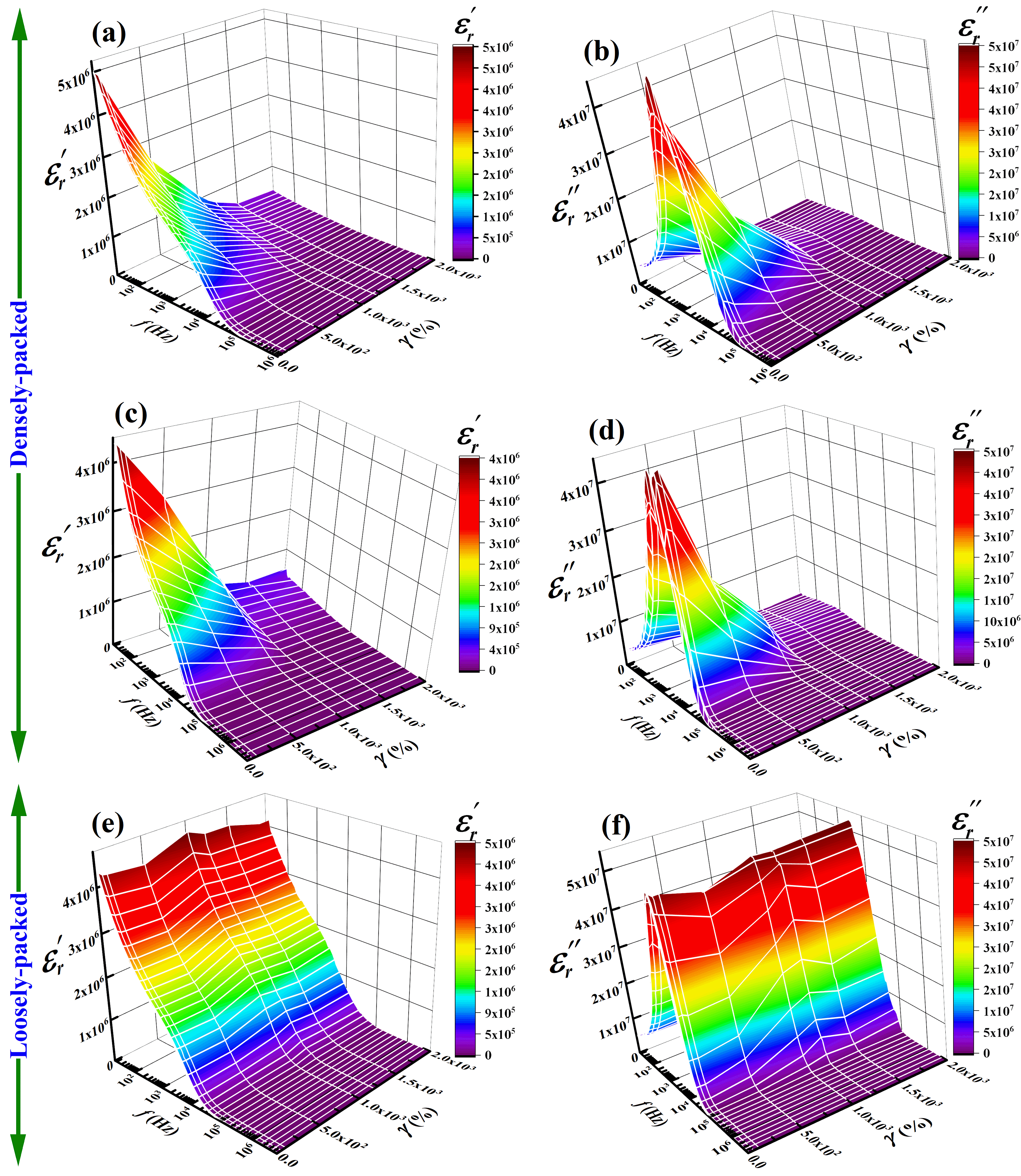}
	\caption{Three dimensional representations at 20$^{\circ}$C (below the LCST) of the (a) real ($\varepsilon_{r}^{'}$) and (b) imaginary ($\varepsilon_{r}^{''}$) parts of the relative dielectric permittivities of aqueous PNIPAM suspensions of effective volume fraction $\phi_{eff}$ = 1.9, (c) $\varepsilon_{r}^{'}$ and (d) $\varepsilon_{r}^{''}$ for effective volume fraction $\phi_{eff}$ = 1.6, and (e) $\varepsilon_{r}^{'}$ and (f) $\varepsilon_{r}^{''}$ for effective volume fraction $\phi_{eff}$ = 1.3, over a range of AC electrical voltage frequencies and applied oscillatory strains. The color bars represent the values of $\varepsilon_{r}^{'}$ and $\varepsilon_{r}^{''}$ with their magnitudes increasing from bottom to top.} 
	\label{Fig.2}
\end{figure}
\section{Results and Discussion}
Fig. 2 shows three-dimensional representations of the real and imaginary parts of the dielectric permittivities of aqueous poly($N$-isopropylacrylamide) (PNIPAM) suspensions at 20$^{\circ}$C (below the LCST) versus frequency of the applied alternating electrical voltage and the applied strain amplitude. At very low strains and at a temperature below the LCST, the elasticities of PNIPAM suspensions of the higher effective volume fractions ($\phi_{eff}$ = 1.9 and 1.6) are comparable to each other, but are very different from the sample of the lowest effective volume fraction ($\phi_{eff}$ = 1.3) used here (Figs. S5(a) of SI). We designate the samples of $\phi_{eff}$ = 1.9 and 1.6 as densely-packed suspensions and the one of $\phi_{eff}$ = 1.3 as a loosely-packed suspension. Figs. S6 and S7 of SI show three-dimensional representations of the dielectric permittivities as functions of strain amplitudes and AC electrical voltage frequencies for the densely-packed and loosely-packed aqueous PNIPAM suspensions at temperatures 33$^{\circ}$C (near the LCST) and 55$^{\circ}$C (above the LCST) respectively. From Figs. 2(a-d), we note that both real and imaginary dielectric permittivities decrease with increasing applied strain amplitudes for the densely-packed PNIPAM suspensions ($\phi_{eff}$ = 1.9 and 1.6) below the LCST. The observed decreases in $\varepsilon_{r}^{'}$ and $\varepsilon_{r}^{''}$ due to an increase in applied strains can be attributed to enhanced structural deformations of the swollen PNIPAM particles in aqueous suspensions at high effective volume fractions. The color bars in Figs. 2, S6 and S7 of SI represent the values of $\varepsilon_{r}^{'}$ and $\varepsilon_{r}^{''}$ with their magnitudes increasing from bottom to top. We observe that for the loosely-packed PNIPAM suspension, the dielectric permittivities are insensitive to the applied strain amplitude at all temperatures (Figs. (e-f) of 2, S6 and S7). For the densely-packed PNIPAM suspensions, in contrast, the real and imaginary permittivities are independent of the applied shear only at temperatures near and above the LCST (Figs. (a-d) of S6 and S7 of SI). For significant modifications in the dielectric properties of homogeneous polymeric systems, the applied deformation must be greater than the equilibrium molecular motion (shear rate > relaxation time$^{-1}$) \cite{HWATANABE1999,watanable2003}. Small deformations can affect the dielectric properties of glassy systems due to the presence of dynamically heterogeneous structures \cite{HWATANABE1999,watanable2003}. Our results indicate that densely-packed PNIPAM suspensions at a temperature below the LCST are possibly in a disordered and metastable (glassy) state constituted by PNIPAM particle assemblies. For the loosely-packed aqueous PNIPAM suspension at all temperatures and the densely-packed suspensions at temperatures near or above the LCST, the observed insensitivity of the dielectric responses to applied deformations presumably arises because of an absence of self-assembled particle aggregates. Smaller particle assemblies have much higher relaxation rates and therefore require much larger shears to induce appreciable changes in the dielectric properties, hence the observed insensitivity of the measured dielectric responses to applied shears in the last scenario.
\begin{figure}[!t]
		\centering
	\includegraphics[width=5.5in,keepaspectratio]{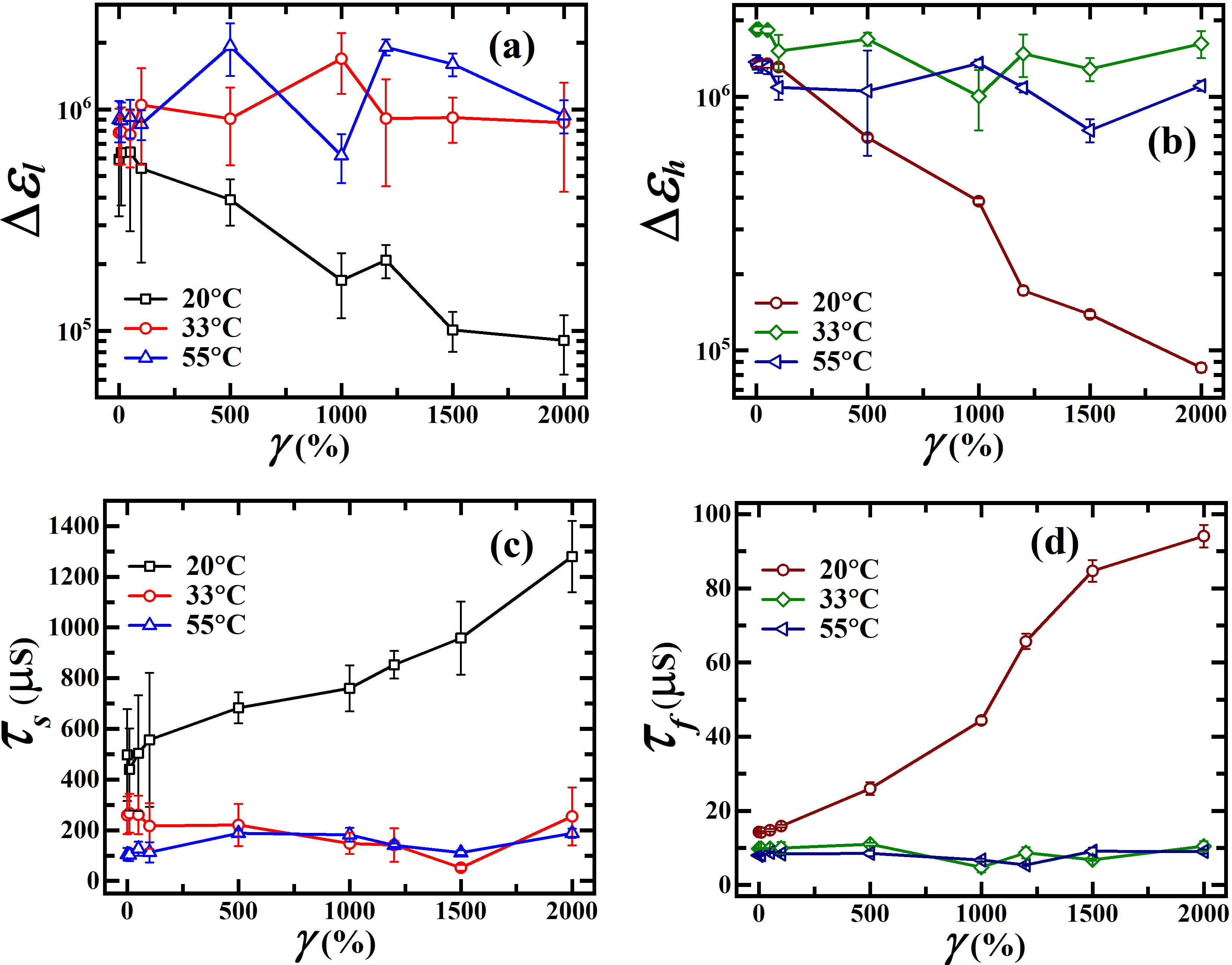}
	\caption{(a) Low frequency ($\Delta\varepsilon_{l}$) and (b) high frequency ($\Delta\varepsilon_{h}$) dielectric relaxation strengths and (c) slow ($\uptau_{s}$) and (d) fast ($\uptau_{f}$) relaxation times as functions of applied oscillatory strain amplitudes for aqueous suspensions of PNIPAM of volume fraction $\phi_{eff}$ = 1.9 at temperatures below, near and above the LCST. These values are obtained by fitting the data in Figs. 2(a), S6(a) and S7(a) to the real part of Eqn. 1.} 
	\label{Fig.3}
\end{figure}
\paragraph{}
The real parts of the dielectric permittivities of aqueous PNIPAM suspensions show two distinct dielectric relaxation processes in the low and high frequency regimes, as can be clearly seen in (a), (c) and (e) of Figs. 2, S6 and S7. We compute the relaxation times and dielectric strengths (determined by the number density of molecules, magnitudes of the moments and orientation fluctuations of the dipoles \cite{CAPACCIOLI20074267,HWATANABE1999}) of all the PNIPAM suspensions prepared at temperatures below, near or above the LCST, by fitting the acquired $\varepsilon_{r}^{'}$ data to the real part of Eqn. 1 (fitted data are shown in Fig. S8 of SI). The dielectric strengths and relaxation times provide important information about the orientation and dynamics of the PNIPAM particles under shear flow. It has been reported that the alignment of dipoles parallel and perpendicular to the electric field respectively increase and decrease the measured dielectric strengths \cite{CAPACCIOLI20074267}. The values of the fitted parameters in the electrode polarization term ($A$ and $m$) of Eqn. 1, which originates from the mobility of the free ions and contributes to the low frequency dielectric permittivities, are listed in Tables 2-4 of SI. In Figs. 3 (a-b) and S9 (a-b) of SI, we show the low- and high frequency dielectric relaxation strengths ($\Delta\varepsilon_{l}$ and $\Delta\varepsilon_{h}$ respectively) of densely-packed PNIPAM suspensions ($\phi_{eff}$ = 1.9 in Figs. 3(a-b) and $\phi_{eff}$ = 1.6 in Figs. S9(a-b) of SI) at all the experimental temperatures. Figs. S10 (a-b) display the values of $\Delta\varepsilon_{l}$ and $\Delta\varepsilon_{h}$ when oscillatory strains were applied to the loosely-packed PNIPAM suspension ($\phi_{eff}$ = 1.3) at the same temperatures. For all the PNIPAM suspensions, $\Delta\varepsilon_{l}$, which is understood to originate from the segmental motion of PNIPAM chains over the entire temperature range as reported in an earlier work \cite{ManYang2017}, shows an enhancement with increasing suspension temperatures. This can be attributed to the increase in polymer density inside the PNIPAM particles due to particle shrinkage/ collapse above the LCST. Interestingly, we note that both $\Delta\varepsilon_{l}$ and $\Delta\varepsilon_{h}$ decrease strongly with increase in applied strain amplitudes for densely-packed PNIPAM suspensions at a temperature below the LCST. It has been proposed that the application of shear strain disrupts the structured polymer particles \cite{CAPACCIOLI20074267} that are present in densely-packed (jammed) suspensions and randomizes the directions of the dipoles. Strong shears are expected to result in the alignment of dipoles parallel to the flow direction. In our experimental geometry, the flow direction is perpendicular to the electric field and results in the observed decrease in net polarization and dielectric strengths. In contrast, no significant strain-induced changes in $\Delta\varepsilon_{l}$ and $\Delta\varepsilon_{h}$ are observed for loosely-packed PNIPAM suspension at all experimental temperatures and for densely-packed PNIPAM suspensions at temperatures near and above the LCST. The strain insensitivity of the dielectric strengths under these conditions arises as the PNIPAM particles do not self-assemble to form strong structures due to low effective volume fractions. We therefore conclude that the packing densities and temperature dependent morphological changes of PNIPAM particles effectively control the dielectric properties and shear responses of their aqueous suspensions.

\paragraph{}
Apart from the dielectric strength, the other parameters that we study to quantify the responses of the charged species are the fast and slow relaxation times \cite{Wenjuan2014,ManYang2017}, extracted by fitting the observed two-step decay of the real part of the dielectric permittivities to Eqn. 1. Figs. 3(c-d) and S9(c-d) display the slow ($\uptau_{s}$) and fast ($\uptau_{f}$) relaxation times of densely-packed aqueous PNIPAM suspensions at $\phi_{eff}$ = 1.9 and 1.6 respectively as a function of strain amplitudes at all the temperatures explored in this study. Figs. S10(c-d) represent the slow and fast relaxation times as a function of strain amplitude for the loosely-packed PNIPAM suspension (i.e. $\phi_{eff}$ = 1.3) at the same temperatures. We see that both $\uptau_{s}$ and $\uptau_{f}$ decrease with increase in suspension temperature for all the samples at a fixed effective volume fraction. We attribute the observed decrease in time scales with increasing temperatures to the reduction in effective volume fractions and enhanced mobilities of the collapsed PNIPAM particles \cite{Manohar_V_2003}. Interestingly, we see from Figs. 3(c-d) and Figs. S9(c-d) of SI that $\uptau_{s}$ and $\uptau_{f}$ increase monotonically with increasing strain amplitudes for the densely-packed PNIPAM suspensions at a temperature below the LCST, thereby demonstrating a slowdown of the polarization fluctuation dynamics with increasing strain amplitude. However, the dielectric relaxation time scales of these samples are insensitive to the applied shear at temperatures near and above the LCST where the particles have already collapsed due to expulsion of water. Furthermore, we note that $\uptau_{s}$ and $\uptau_{f}$ are relatively unaffected by the applied strain amplitudes at all temperatures when PNIPAM particles are loosely-packed in aqueous suspension (Figs. S10(c-d) of SI). The time scales extracted from our rheo-dielectric measurements therefore reveal the distinct dynamical properties of PNIPAM particles in aqueous suspensions at different temperatures and under different externally imposed mechanical stresses. 

\paragraph{}
Moreover, from Figs. 2, S6 and S7, we observe that the imaginary contributions to the dielectric permittivities show a single loss peak. Figs. S11(a-c) of SI show the relaxation times estimated from the peak frequency values of the dielectric losses ($\uptau = 1/2\pi f_{peak}$), where $f_{peak}$ is the frequency at which the dielectric loss peak is noted. We see that the relaxation times $\uptau$ are of the order of the fast relaxation processes estimated from fits of the real part of the dielectric permittivities to the Cole-Davidson relaxation model (Figs. 3(d) and S9(d) of SI). We therefore conclude that the dominant relaxation mechanism responsible for the observed dielectric loss originates from counterion polarization below the LCST. In contrast, at temperatures above the LCST, the dielectric loss can be attributed to interfacial polarization generated by the free charged species present in the suspension and the dynamics of counterions around the particles. In results that are consistent with data displayed in Figs. 3(d) and S9(d), we note from Figs. S11(a-b) that the dielectric relaxation times, $\uptau$, of densely-packed PNIPAM suspensions ($\phi_{eff}$ = 1.9 and 1.6) increase with increase in applied strain amplitudes at a temperature below the LCST. However, the relaxation rates are independent of temperature for densely-packed PNIPAM suspensions at temperatures near and above the LCST and for the loosely-packed suspension ($\phi_{eff}$ = 1.3) at all temperatures. 
\begin{figure}[!t]
	\centering
	\includegraphics[width=2.5in,keepaspectratio]{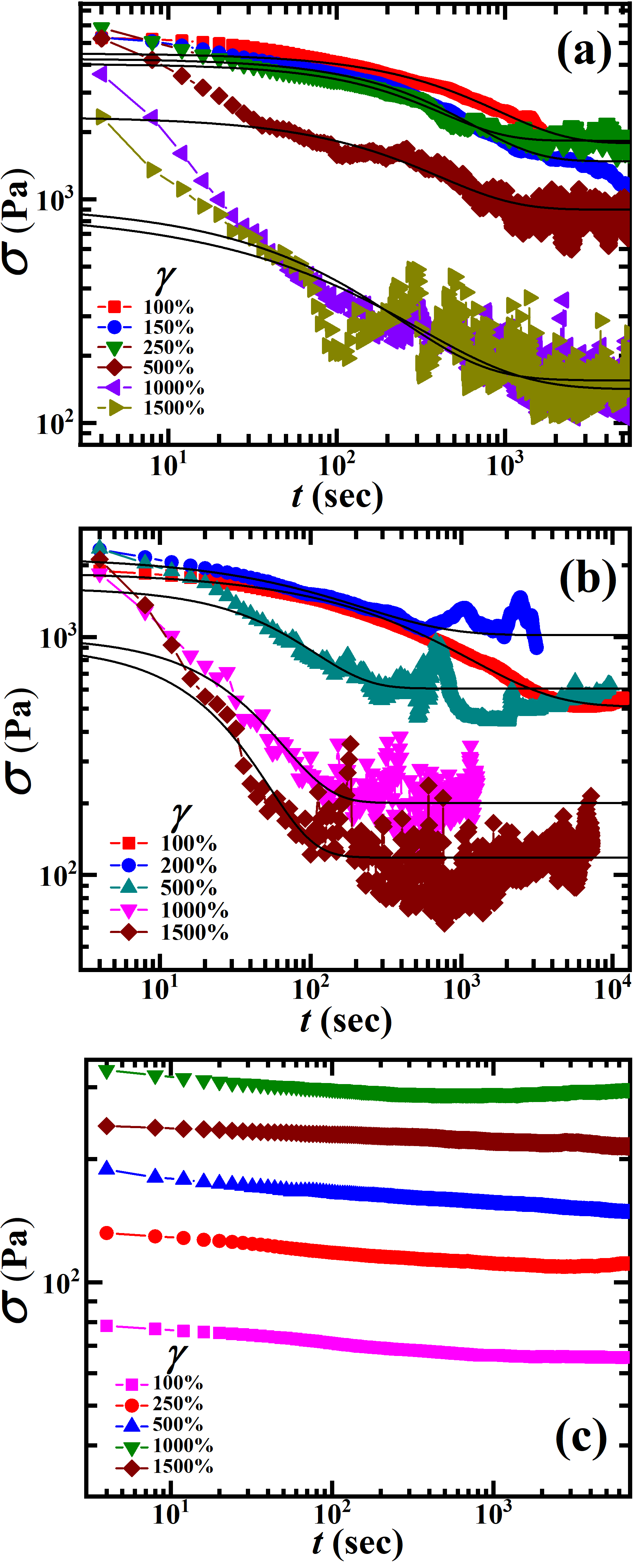}
	\caption{Oscillatory shear stress response {\it vs.} time exhibited by PNIPAM suspensions of effective volume fractions $\phi_{eff}$ = (a) 1.9, (b) 1.6 and (c) 1.3 at temperature 20$^{\circ}$C (below the LCST) and for different applied oscillatory strains $\gamma$.} 
	\label{Fig.4}
\end{figure}
\begin{figure}[!t]
	\centering
	\includegraphics[width=5.0in,keepaspectratio]{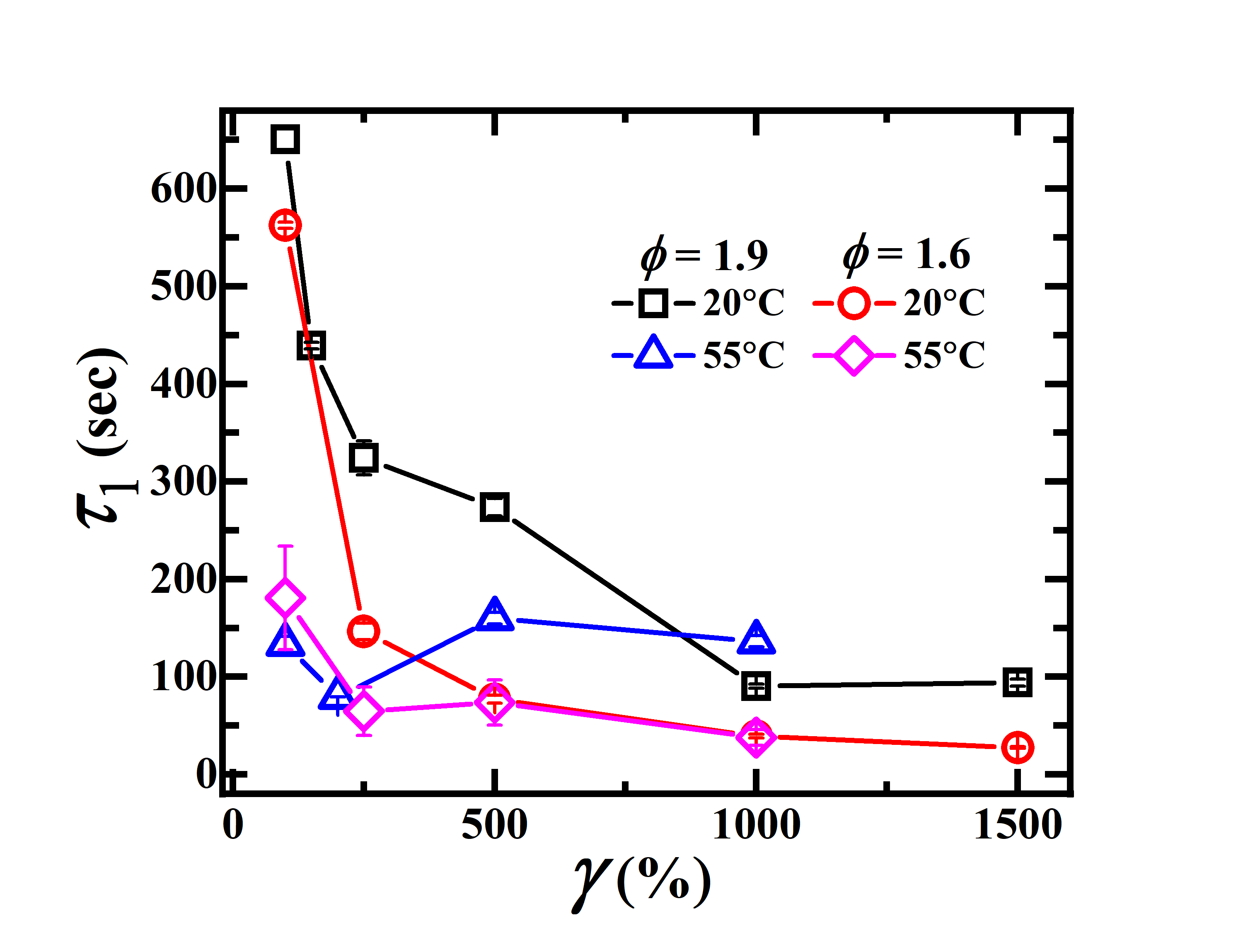}
	\caption{Stress relaxation times, $\uptau_{1}$, of densely-packed aqueous suspensions of PNIPAM particles at temperatures 20$^{\circ}$C and 55$^{\circ}$C as a function of applied oscillatory strain amplitude.} 
	\label{Fig.4}
\end{figure}
\paragraph{}
To complement our rheo-dielectric study that uncovers counter-intuitive polarization fluctuation dynamics upon applying oscillatory shear deformations, we next study the influence of applied large amplitude oscillatory strains on the bulk mechanical moduli of PNIPAM suspensions by performing independent stress relaxation experiments. In these measurements, the range of the large amplitude oscillatory strains is comparable to that applied in our rheo-dielectric experiments. Our experimental design therefore allows us to study the responses of PNIPAM suspensions over a length scale range varying over several (7-8) orders of magnitude. Stress relaxation decays, exhibited by densely-packed (i.e. $\phi_{eff}$ = 1.9 and 1.6) and loosely-packed (i.e. $\phi_{eff}$ = 1.3) aqueous PNIPAM suspensions at temperatures below, near and above the LCST, upon the application of oscillatory strains of pre-determined amplitudes, are shown in Figs. 4, S12, and S13, respectively. For the densely-packed PNIPAM suspensions at/near the LCST, power-law stress relaxations are observed (Fig. S12(a-b)). For the loosely-packed suspension under the same condition, the stress decay flattens out very rapidly (Fig. S12(c)). Interestingly, the two-step stress relaxation processes observed for the densely-packed PNIPAM suspensions (Figs. 4(a-b) and S13(a-b)) below and above the LCST are reminiscent of the relaxation of glassy materials or entangled polymeric melts \cite{Ranjini_2010,D_Feldman}. Given the extremely rapid decay of the short-time stress relaxation, we could not acquire adequate reliable data to model the fast stress relaxation process satisfactorily. We therefore focus only on the slow bulk stress relaxation data which is well described by the Kohlrausch–Williams–Watts (KWW) model \cite{Williams_1970}. The KWW model represents the time dependent stress relaxation of the slower mode in glassy materials as a stretched exponential function: 
\begin{equation}
    \sigma(\uptau) = A exp[-(\uptau/\uptau_{1})^{\beta}]
\end{equation}
where, $\sigma(\uptau)$, $A$, $\uptau{_1}$ and $\beta$ are the time dependent shear stress, intercept of the slow relaxation decay, time scale corresponding to the slow mode of the stress relaxation process and stretching exponent (0 < $\beta$ < 1) respectively. We note that the stress relaxation times for densely-packed PNIPAM suspensions, which are obtained by fitting the data plotted in Figs. 4(a-b) to Eqn. 2, decrease with applied shear strain amplitude below the LCST (Fig. 5), thereby revealing the presence of bulk shear thinning behavior at low temperatures. Interestingly, therefore, while the polarization fluctuation in a suspension of densely-packed PNIPAM particles slows down with an increase in the applied strain below the LCST, the bulk stress relaxation speeds up simultaneously. 

	\begin{figure}[!t]
	\centering
	\includegraphics[width=5.5in,keepaspectratio]{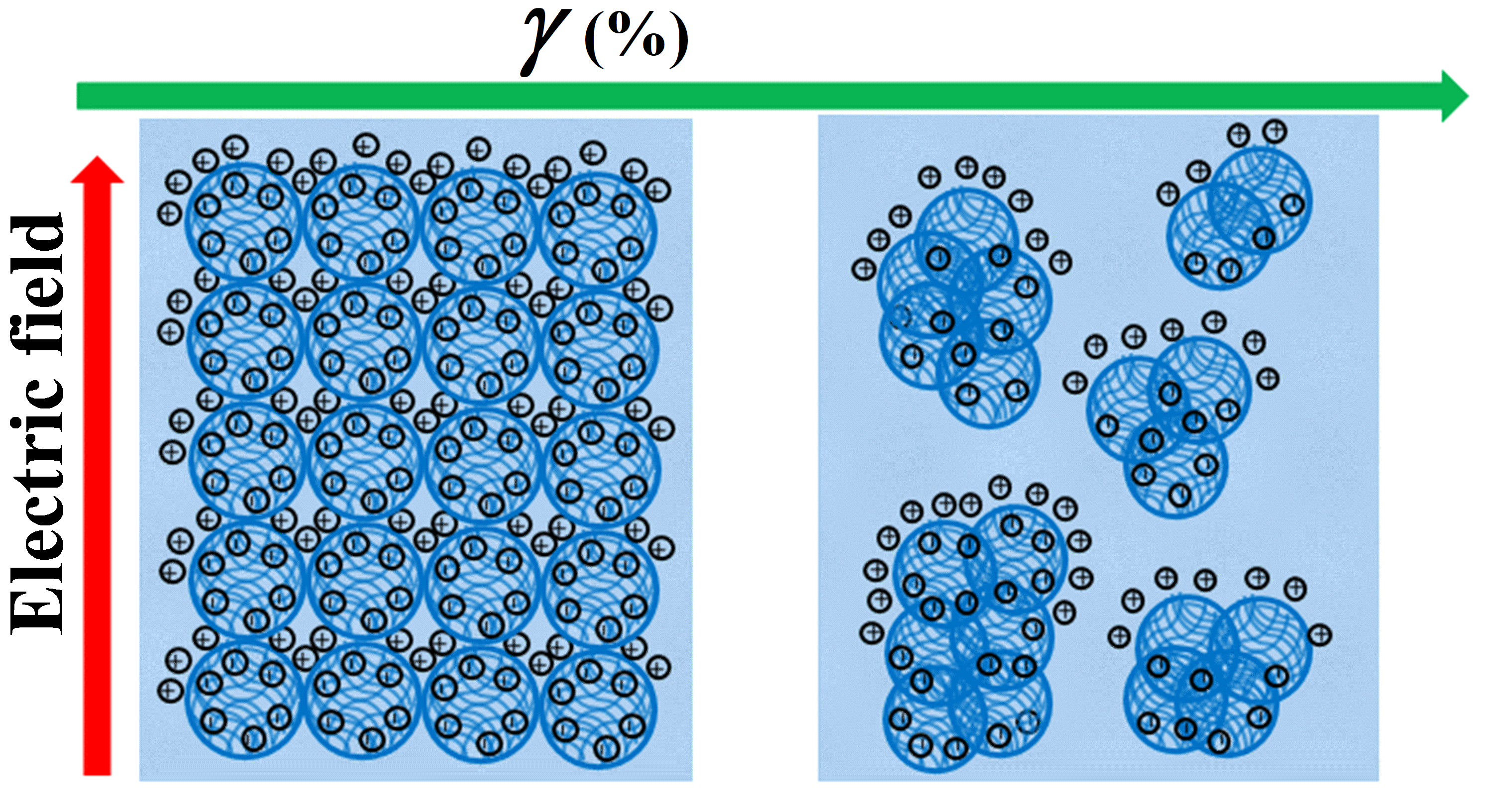}
	\caption{A schematic illustration of a densely-packed PNIPAM suspension under an applied strain. The green arrow at the top and the red arrow on the left represent, respectively, the increase in applied strain amplitude and the direction of the electric field.} 
	\label{Fig.4}
\end{figure}

The observed dichotomous dynamics at macroscopic and nanometre length scales can be understood by considering the shear-induced formation and rupture of fragile PNIPAM particle clusters in densely-packed suspensions. It has been reported in earlier work that PNIPAM particles in dense aqueous suspensions phase separate by forming clusters under shear flow \cite{Manohar_V_2003,Stieger2003-us}. We attribute the slowing down of the slow and fast dielectric relaxation processes to the formation of such microscopic PNIPAM particle clusters. This is illustrated in Fig. 6 which schematically represents polarization fluctuations in the sample (left-panel) and breakup of PNIPAM clusters (right-panel) due to stresses imposed by electric fields and shear strains respectively. Badiger et al. \cite{Manohar_V_2003} have observed stress overshoots in aqueous suspensions of PNIPAM particles undergoing shear flow due to the presence of entangled PNIPAM polymers. Such shear-induced entanglements of PNIPAM chains in densely-packed suspensions below the LCST are expected to restrict the segmental motion of PNIPAM chains, resulting in the observed increase of the slow dielectric relaxation times. Furthermore, counterion separation is enhanced due to the fragmentation of particle clusters undergoing shear deformation (Fig. 6, right-hand panel). This enhances the particle dipole moments and causes the dielectric fast relaxation processes to slow down. The simultaneous decrease in the dielectric strengths with increasing shears arises from the randomization of the dipoles and the distortions and breakups of PNIPAM clusters by the applied shear. Furthermore, the decrease in the bulk viscosity of PNIPAM suspensions due to shear induced breakdown of the fragile clusters results in accelerated particle dynamics and shear thinning at macroscopic length scales.

\section{Conclusions}
The bulk mechanical response and the polarization fluctuations of sheared complex fluids have independently been studied in the literature using rheometry and dielectric spectroscopy \cite{Senff1999,CMisra2020,J_Wu_2003,Wu_2003_PRL,Romeo2010,Wenjuan2014,ManYang2017,J_Zhou2012}. However, to the best of our knowledge, simultaneous rheological and dielectric experiments have never been performed to study the dynamics of sheared colloidal suspensions over spatiotemporal scales spanning several decades. Our work attempts to bridge this gap by simultaneously studying the dielectric and stress relaxation behaviors of aqueous suspensions of colloidal hydrogel poly($N$-isopropylacrylamide) (PNIPAM) particles under large applied oscillatory shear strains. We observe two distinct dielectric relaxation processes in the low and high frequency regimes. The low frequency relaxation process is understood to originate from the segmental motion of the polymer and the motion of counterions along the polymers chains \cite{ManYang2017} over the entire temperature range. On the other hand, the high frequency relaxation is believed to arise from counterion fluctuations at a temperature below the LCST and from interfacial polarization above the LCST. We propose that the observed decrease of the dielectric relaxation strengths in densely-packed PNIPAM suspensions upon the application of oscillatory shear strains at a temperature below the LCST is due to the distortion of dynamically heterogeneous suspension structures and randomization in the orientations of the constituent dipoles. Notably, the dielectric responses of loosely-packed PNIPAM suspensions at all temperatures and densely-packed PNIPAM suspensions at temperatures near and above the LCST are insensitive to the applied deformations. The most significant finding of this study is that while the polarization fluctuations of densely-packed PNIPAM suspensions slow down with increase in applied oscillatory strain amplitudes at a temperature below the LCST, the bulk stress relaxation dynamics speed up under identical conditions. Earlier studies such as those by Su et al. \cite{Wenjuan2014} and Yang et al. \cite{ManYang2017} reported the dielectric behavior of relatively dilute aqueous PNIPAM suspensions (0.005 wt.\% and 5.6 wt.\% respectively). The dielectric strengths and time scales reported in the present study are therefore much higher than for the relaxation dynamics reported earlier \cite{Wenjuan2014,ManYang2017}. The dense PNIPAM suspensions in our experiments are characterised by a significantly enhanced number of dipoles that restrict the segmental motion of PNIPAM chains, resulting in the novel dynamics reported in our study.

\paragraph{}
Besides revealing the rich dynamical behavior of dense suspensions of PNIPAM hydrogels at different length scales, we show here that the self-assembly and rheological properties of these samples can be fine-tuned by controlling temperature and the applied strain deformation. These hydrogels are therefore attractive candidates for engineering energy storage devices, such as stretchable, soft and flow batteries \cite{Helal_2016} and flow capacitors \cite{Presser_2012}. We strongly believe that the present study can serve as a starting point for further research into the length scale dependent structural and dynamical properties of dense colloidal (soft glassy) suspensions and other kinetically constrained complex systems.
\section*{Acknowledgments}
The authors thank K. N. Vasudha for help with DSC measurements and N. V. Madhusudana for useful discussions. We acknowledge Department of Science and Technology, Science and Engineering Research Board for funding our research (grant number EMR/2016/006757).

 \bibliographystyle{ieeetr}
\bibliography{Ref_Rheo_di.bbl}

\end{document}